\documentstyle[12pt,epsfig]{article}
\def\lsim{\:\raisebox{-0.5ex}{$\stackrel{\textstyle<}{\sim}$}\:}

\begin{document}
\begin{flushright}
BI-TP 97/31
\end{flushright}

\vskip 45pt
\begin{center}
{\Large \bf On $e^+e^- \to W^+W^-$ at LEP2${}^{\displaystyle \ast}$}

\vspace{11mm}

{\large D. Schildknecht}

\vspace{150pt}

{\bf ABSTRACT}
\end{center}
\begin{quotation}
We briefly discuss the first {\it direct} experimental evidence from 
LEP2 on non-vanishing trilinear couplings among the weak vector bosons.
Subsequently we review the improved Born approximation to include
radiative corrections. Provided the appropriate high-energy scale is used
for the SU(2) gauge-coupling, the improved Born approximation is
sufficiently accurate for all practical purposes at LEP2.
\end{quotation}

\vspace*{\fill}
\footnoterule
{\footnotesize
\noindent ${}^{\displaystyle \ast}$
Supported by
EU-Project CHRX-CT 94-0579. To appear in the Proceedings of the EU-network
meeting, Ouranoupolis, Greece, April 1997, edited by A. Nicolaidis}

\newpage
The process of W-pair production in $e^+e^-$ annihilation is presently
studied experimentally at LEP2. In the future, it will be one of the
outstanding processes at a linear collider in the TeV energy range. The
process of W-pair production
yields direct experimental 
\begin{figure}[htpb] 
\unitlength 1mm
\begin{picture}(152,152)
\put(0,0){
\epsfxsize=15cm
\epsfysize=16cm
\epsfbox{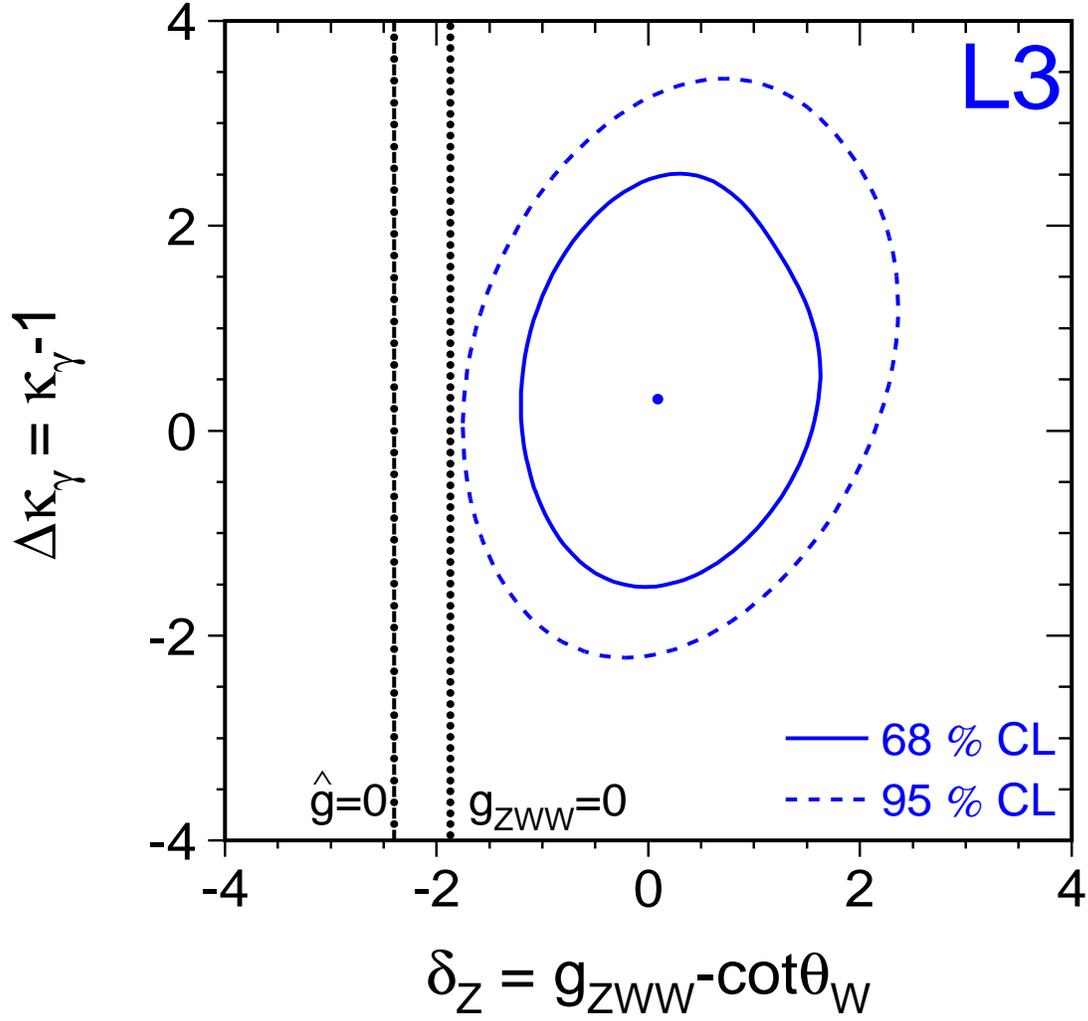}
}
\end{picture}
\caption{\it Contour curves of 68 \% and 95 \% confidence level in the
$(\delta_Z, \Delta \kappa_\gamma)$ plane are shown as solid and dashed
lines. Expectations due to vanishing ZWW and weak couplings, $g_{ZWW} = 0$
and $\hat g = 0$, are indicated by the dotted and dashed-dotted lines.
(From [3]).}
\end{figure}
information on the non-Abelian 
couplings characteristic for the $SU(2) \times U(1)$ electroweak
theory and allows to put bounds on potential non-standard $Z_0W^+W^-$
and $\gamma W^+W^-$ couplings [1,2]. Indeed, even after a fairly short
time of running, LEP2 results from the L3 collaboration have provided
direct experimental evidence [3] for the existence of a genuine trilinear
coupling between the members of the $W^\pm, W^0$ triplet characteristic
for the non-Abelian structure of the basic Lagrangian. A vanishing
genuine trilinear coupling in the basic electroweak Lagrangian, $\hat g =
0$, is excluded at 95 \% C.L. in a two-parameter, $\hat g, \Delta \kappa_\gamma
\equiv \kappa_\gamma - 1$, analysis of the experimental data, where
$\kappa_\gamma$ denotes the electromagnetic dipole coupling of the $W^\pm$.
This is shown
in fig. 1 taken from ref. [3]. 
The analysis of the experimental data,
without much loss of generality, is based on custodial SU(2) symmetry
for the couplings among the vector bosons, i.e. on 
imposing restoration of SU(2) for vanishing electromagnetic coupling,
$e = 0$ [4,2]. Under this constraint, deviations from the standard value,
of the $Z_{\mu\nu}$ dipole coupling, $\kappa_Z = 1$, 
are proportional to $\Delta \kappa_\gamma$,
thus reducing the number of independent dim.4 couplings in the 
phenomenological vector-boson Lagrangian from three to two, the $Z^0W^+W^-$
coupling, $g_{ZW^+W^-}$, and the electromagnetic dipole coupling
$\kappa_\gamma$ of the $W^\pm$. In fig. 1, the experimental bounds on the
deviations of these couplings from their Standard Model values, $\delta_Z
\equiv g_{ZWW} - \cot \theta_W$ and $\Delta \kappa_\gamma \equiv 
\kappa_\gamma -1$, are shown.

In what follows, I will concentrate on briefly describing a simple
approximate treatment of radiative correction to $e^+e^- \to W^+W^-$ relevant
at LEP2 energies. It is based on a recent paper in collaboration with
Masaaki Kuroda and Ingolf Kuss [5], which in turn rests heavily on previous
work by B{\"o}hm, Denner and Dittmaier [6]. Essentially, I wish to stress
that an accuracy of the order of 0.5 \% can be reached for the
radiatively corrected differential cross section of $e^+e^- \to W^+W^-$ at
LEP2 energies by supplementing the Born approximation by Coulomb
corrections and initial-state-radiation. An appropriate choice of the
scale, in particular for the SU(2) gauge coupling, and also for the 
electromagnetic coupling in the
Born approximation is a prerequisite for obtaining this accuracy of 
approximately 0.5 \%.

The Born amplitude is most transparently derived [5] directly from the basic
Lagrangian of the $SU(2) \times U(1)$ theory by treating  $BW_3$ mixing
perturbatively (to all orders of the mixing). Compare the Feynman diagrams
in fig. 2. 
\begin{figure}[htpb] 
\unitlength 1mm
\begin{picture}(152,152)
\put(0,0){
\epsfxsize=15cm
\epsfysize=16cm
\epsfbox{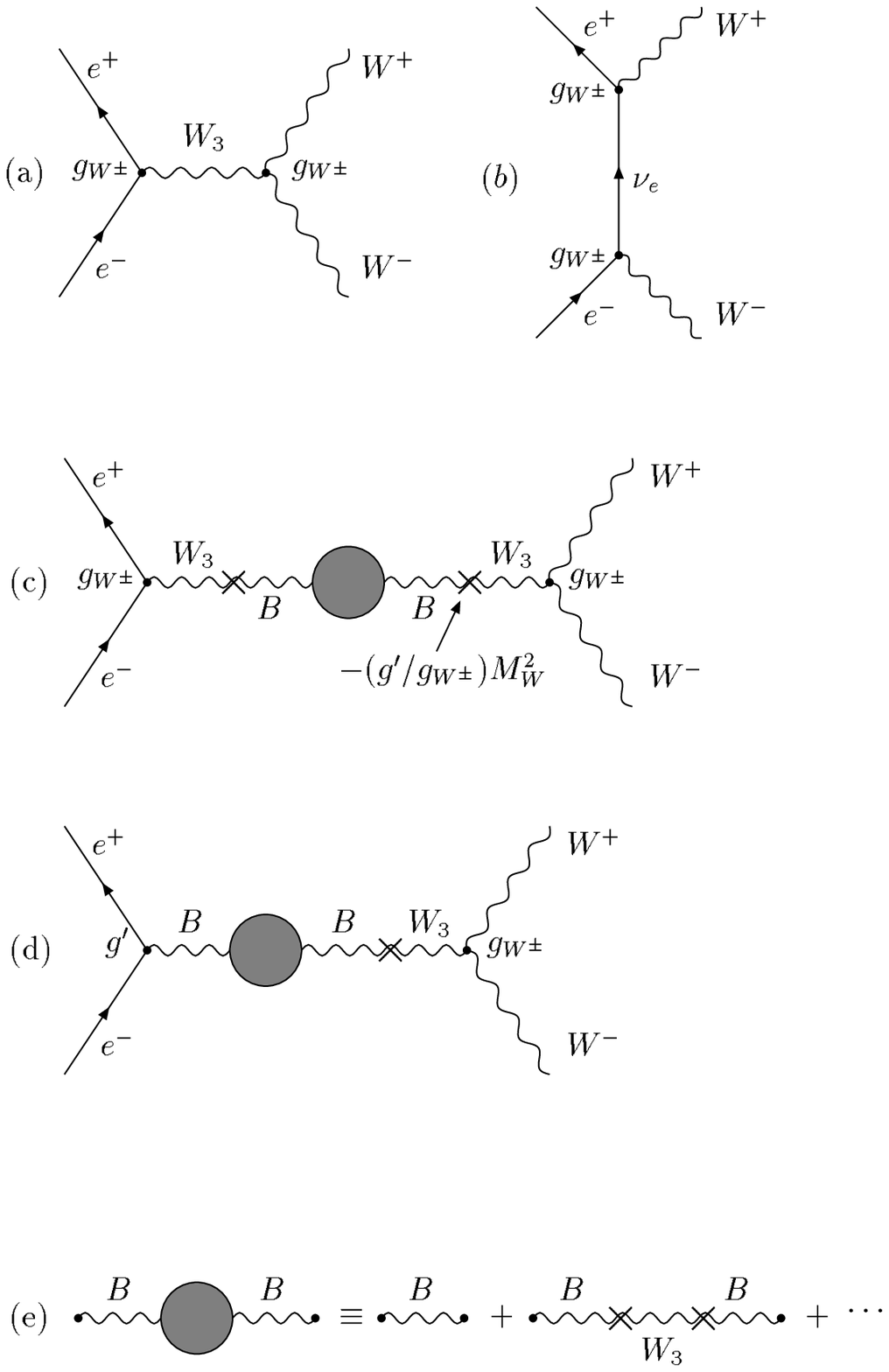}
}
\end{picture}
\caption{\it Evaluating the electroweak Born approximation in the 
$BW_3$ basis.  }
\end{figure}
Upon replacing $g^{\prime 2}$ by $e^2$ via $g^{\prime 2} \equiv
e^2/(1 - e/g^2_{W^\pm})$, from the Feynman diagrams of fig. 2, 
one obtains the Born amplitude directly in terms of the SU(2) coupling,
$g_{W^\pm}$, and the electromagnetic coupling, $e$,

\begin{equation}
{\cal M}_{Born} (\kappa,\lambda_+,\lambda_-,s,t) = g^2_{W^\pm} 
{1 \over 2} \delta_{\kappa^-} {\cal M}_I
+ e^2 {\cal M}_Q, \label{one}
\end{equation}
\newpage\noindent
with the weak isospin amplitude (modified by the replacement of
$M^2_{W^\pm} \to M^2_Z = (1 + g^{\prime 2}/g^2_{W^\pm}) M^2_W$ 
in the propagator via $BW^3$ mixing) being given
by
\begin{equation}
{\cal M}_I = {1 \over {s-M^2_Z}} {\cal M}_s + {1 \over t} {\cal M}_t,
\label{two}
\end{equation}
while the electromagnetic amplitude has the double-pole structure
\begin{equation}
{\cal M}_Q = -{M_Z^2 \over s(s-M_Z^2)} {\cal M}_s.
\label{three}
\end{equation}
For the explicit form of the s-channel and t-channel quantities, ${\cal M}_s$
and ${\cal M}_t$ in terms of the kinematic variables $s$ and $t$ and the
particle helicities we refer to ref. [7].

The calculation of the cross section for $e^+ e^- \to W^+ W^-$ from (1)
requires a specification of the scale at which the SU(2) gauge coupling,
$g_{W^\pm}$, and the electromagnetic coupling, $e$, in (1) are to be
defined. As $W$-pairs are produced at energies of $2M_{W^\pm} \lsim
\sqrt s \sim 200 GeV$ at LEP2, it is natural to choose a high-energy
scale, such as $\sqrt s$. We expect it to be sufficiently accurate to use
the scale $M_W \cong M_Z$ instead of $\sqrt s$. Accordingly, we define
$g_{W^\pm} (M^2_W)$ by the leptonic branching ratio of the 
$W^\pm$ boson, $\Gamma^W_l$ [8],
\begin{equation}
g^2_{W^\pm} (M^2_{W^\pm}) = 48 \pi {{\Gamma_l^W}\over {M_{W^\pm}}}.
\label{four}
\end{equation}
The $W$ branching ratio not being experimentally known with sufficient
accuracy, it must be expressed in terms of the Fermi coupling, $G_{\mu}$.
The $SU(2)$ gauge coupling (\ref{four}) then becomes [8]
\begin{equation}
g^2_{W^\pm} (M^2_{W^\pm}) = {{4 \sqrt 2 G_\mu M^2_{W^\pm}} \over {1 + 
{\Delta y}^{SC}}},
\label{five}
\end{equation}
where the one-loop correction, $\Delta y^{SC}$, derives from the 
change of scale from $\mu^{\pm}$ decay, where $G_\mu$ is defined, to
$W^\pm$ decay. It (obviously) contains a sum of two contributions,
\begin{equation}
{\Delta y}^{SC} = \Delta y^{SC}_{ferm} + {\Delta y}^{SC}_{bos}.
\label{six}
\end{equation} 
The fermion contribution, essentially due to contributions of the light
leptons and quarks to the $W$ propagator, leads to an increase of $g_{W^\pm}$,
when taken by itself, as
\begin{equation}
\Delta y^{SC}_{ferm} \vert_{m_t = 180 GeV} = - 7.79 
\times 10^{-3}.\footnote{The dependence on the mass of the top quark, $m_t$,
is practically negligible, as $\Delta y^{SC}_{ferm} \vert_{m_t \to \infty} 
= - {{3 \alpha (M^2_Z)} \over{4 \pi s^2_0}} \simeq - 8.01 \cdot 10^{-3}$.}
\label{seven}
\end{equation}
The bosonic contribution, on the other hand, with
\begin{equation}
{\Delta y}^{SC}_{bos} = 11.1 \times 10^{-3},
\label{eight}
\end{equation}
leads to a decrease of $g_{W^\pm}$.
The overall correction,
\begin{equation}
{\Delta y}^{SC} = 3.3 \times 10^{-3},
\label{nine}
\end{equation}
which is practically independent of the top quark and Higgs boson masses,
thus results in a decrease of the coupling $g_{W^\pm} (M^2_{W^\pm})$
in (\ref{five}) relative to the
low-energy value, $g^2_{W^\pm} (0) \equiv 4 \sqrt 2 G_\mu
M^2_{W^\pm}$. Finally, we have to specify the electromagnetic coupling,
$e$, in (\ref{one}), which is given by [9]
\begin{equation}
\left({{e^2} \over {4 \pi}}\right)^{-1} = 
\alpha^{-1} (M^2_Z) = 128.89 \pm 0.09.
\label{ten}
\end{equation}

\begin{table}
\begin{center}
\vspace{-5mm}
\begin{tabular}{|r|r|r|c||r|r|c|}
\hline
\multicolumn{1}{|c|}{angle}&\multicolumn{3}{c||}{unpolarized}&
\multicolumn{3}{c|}{left-handed}\\
\hline
&\multicolumn{1}{c|}{$\Delta_{IBA}$}&$\delta\Delta_{IBA}$&
\multicolumn{1}{c|}{$\Delta_{
IBA}+\delta\Delta_{IBA}$}&
$\Delta_{IBA}$&$\delta\Delta_{IBA}$&
\multicolumn{1}{c|}{$\Delta_{IBA}+\delta\Delta_{IBA}$}\\
\hline\hline
\multicolumn{7}{|c|}{$\sqrt{s}=161$ GeV}\\
\hline
\multicolumn{1}{|c|}{total}&1.45&-0.72&0.73&1.45&-0.72&0.73\\
10&1.63&-0.73&0.90&1.63&-0.73&0.90\\
90&1.44&-0.72&0.72&1.44&-0.72&0.72\\
170&1.26&-0.70&0.56&1.26&-0.70&0.56\\
\hline\hline
\multicolumn{7}{|c|}{$\sqrt{s}=165$ GeV}\\
\hline
\multicolumn{1}{|c|}{total}&1.27&-0.71&0.56&1.28&-0.71&0.57\\
10&1.67&-0.74&0.93&1.67&-0.74&0.93\\
90&1.17&-0.71&0.46&1.18&-0.71&0.47\\
170&0.75&-0.67&0.08&0.77&-0.67&0.10\\
\hline\hline
\multicolumn{7}{|c|}{$\sqrt{s}=175$ GeV}\\
\hline
\multicolumn{1}{|c|}{total}&1.26&-0.71&0.55&1.28&-0.71&0.57\\
10&1.71&-0.75&0.96&1.71&-0.75&0.96\\
90&1.03&-0.69&0.34&1.06&-0.70&0.36\\
170&0.59&-0.62&-0.03&0.69&-0.63&0.06\\
\hline\hline
\multicolumn{7}{|c|}{$\sqrt{s}=184$ GeV}\\
\hline
\multicolumn{1}{|c|}{total}&1.02&-0.70&0.32&1.06&-0.71&0.35\\
10&1.57&-0.75&0.82&1.57&-0.75&0.82\\
90&0.67&-0.68&-0.01&0.72&-0.69&0.03\\
170&0.10&-0.58&-0.48&0.32&-0.64&-0.32\\
\hline\hline
\multicolumn{7}{|c|}{$\sqrt{s}=190$ GeV}\\
\hline
\multicolumn{1}{|c|}{total}&1.24&-0.70&0.54&1.28&-0.71&0.57\\
10&1.67&-0.74&0.93&1.67&-0.75&0.92\\
90&0.95&-0.68&0.27&1.01&-0.69&0.32\\
170&0.58&-0.57&0.01&0.83&-0.59&0.24\\
\hline\hline
\multicolumn{7}{|c|}{$\sqrt{s}=205$ GeV}\\
\hline
\multicolumn{1}{|c|}{total}&1.60&-0.70&0.90&1.65&-0.71&0.94\\
10&1.77&-0.74&1.03&1.77&-0.74&1.03\\
90&1.55&-0.66&0.89&1.64&-0.68&0.96\\
170&1.61&-0.53&1.08&1.94&-0.56&1.38\\
\hline
\end{tabular}
\end{center}

\caption{
  The Table shows the quality of the improved Born approximation (IBA) for the
  total (defined by integrating over $10^0\protect\lsim\vartheta\protect\lsim 
  170^0$) and the
  differential cross section (for $W^-$-production angles $\vartheta$ of
  $10^0,90^0$ and $170^0$) for $e^+e^-\protect\to W^+W^-$ 
  at various energies for
  unpolarized and left-handed electrons. The final percentage deviation,
  $\Delta_{IBA} + \delta \Delta_{IBA}$, of the IBA from the full one-loop 
  result
  is obtained by adding the correction $\delta \Delta_{IBA}$ resulting from
  using the appropriate high energy scale in the SU(2) coupling, to the
  percentage deviation, $\Delta_{IBA}$, based on using the low-energy scale in
  the SU(2) coupling, i.e. $\Delta y^{SC} =0$. (From [5])}

\label{table1}
\end{table}

Improving the Born approximation (\ref{one}) by Coulomb corrections and
initial state radiation [6], the differential cross section for $W^\pm$
pair production in $e^+e^-$ annihilation becomes [5]
\begin{eqnarray}
\left({{d \sigma} \over {d \Omega}}\right)_{IBA} &=& {\beta \over {64 \pi^2 s}}
\left\vert {{2 \sqrt 2 G_\mu M^2_W} \over {1 + {\Delta y}^{SC}}}
{\cal M}^\kappa_I \delta_{\kappa^-} + 4 \pi\alpha (M^2_Z) {\cal M}^\kappa_Q 
\right\vert^2\cr
&&+\left({{d\sigma} \over {d \Omega}}\right)_{Coul} (1 - \beta^2)^2 + 
\left({{d\sigma} \over
{d\Omega}}\right)_{ISR}.\label{11}
\end{eqnarray}
We note that the weak interaction term in (\ref{11}) is dominant relative
to the electromagnetic one. 
Accordingly, the effect of chosing the appropriate high-energy scale
in $g_{W^\pm}$ is most important and may easily be estimated. Neglecting
${\cal M}_Q$, the decrease of the normalized cross section due to
$\Delta y^{SC} \not= 0$,
\begin{equation}
\delta \Delta_{IBA} = {{({{d\sigma} \over {d\Omega}})_{IBA~({\Delta y}^
{SC} \not= 0)} - ({{d\sigma} \over {d\Omega}})_{IBA~({\Delta y}^
{SC} = 0)}}\over {({{d\sigma} \over {d\Omega}})_{Born}}},
\label{12}
\end{equation}
becomes
\begin{equation}
\delta \Delta_{IBA} \simeq - 2 {\Delta y}^{SC} = -0.66\% .
\label{13}
\end{equation}
The quality of the improved Born approximation (\ref{11}) is obtained by
adding $\delta \Delta_{IBA}$ to the quantity $\Delta_{IBA}$, which results
from comparing (\ref{11}) for $\Delta y^{SC} = 0$ with the full one-loop
results [6,7,10]. 
The detailed numerical results
results in Table 1 [5] indeed show that the improved Born
approximation with inclusion of $\Delta y^{SC} \not= 0$ yields an 
accuracy below 1 \%, which is sufficiently accurate for all practical
purposes.

We finally comment on the significance of the appropriate choice of
the high-energy scale in the weak coupling, $g_W^\pm (M^2_W)$, with 
respect to recent one-loop calculations [11] which incorporate the decay
of the $W^\pm$ into 4 fermions in a gauge-invariant formulation.
These calculations take into account fermion-loops only. While
interesting as a first step towards a full one-loop evaluation
of $e^+ e^- \to 4$ fermions, the numerical results of a calculation
including fermion loops only can easily be estimated within the
present framework of stable $W^\pm$ to enlarge the cross section
appreciably. In fact, taking into account
fermion loops only, the estimate (13) changes sign and becomes
\begin{equation}
\delta\Delta_{IBA}\vert_{ferm}\simeq -2\Delta y^{SC}_{ferm} 
\vert_{m_t = 180 GeV} \simeq +1.56\%, 
\label{14}
\end{equation}
and the total deviation from the full one-loop results (using
$\Delta_{IBA} \simeq 1.2 \%$ from Table 1) rises to values of
\begin{equation}
\Delta_{IBA} + \delta \Delta_{IBA}\vert_{ferm}\cong 2.8 \% .
\label{15}
\end{equation}
Accordingly, results from fermion-loop calculations including the
decay of the $W^\pm$ are expected to overestimate the cross section 
by almost 3 \%
relative to the (so far unknown) outcome of a calculation of 
$e^+e^- \to 4$ fermions including bosonic loops as well. It is
gratifying, that a simple procedure immediately suggests itself for
improving the large discrepancy (15). One simply has to approximate
the bosonic loop corrections by using the substitution 
\begin{equation}
G_\mu\to G_\mu/(1+\Delta y^{SC}_{bos})
\label{16}
\end{equation}
with $\Delta y^{SC}_{bos} = 11.1 \times 10^{-3}$ in the four-fermion
production amplitudes. Substitution (16) practically amounts to
using $g_{W^\pm} (M^2_W)$ in four-fermion production as well.
With substitution (16), it is indeed to be expected that the 
deviation of four-fermion production in the fermion-loop scheme
will be diminished from the above estimated value of $\simeq 2.8 \%$
to a value below 1 \%.

In conclusion, evaluating the Born approximation for $e^+e^- \to W^+W^-$
with the SU(2) gauge coupling and the electromagnetic coupling at the
appropriate high-energy scale, and supplementing with Coulomb corrections
and initial state radiation, yields differential cross sections which are
sufficiently accurate for all practical purposes at LEP2.

Calculations of $e^+ e^- \to $ 4 fermions in the fermion-loop scheme
overestimate the true cross section by a non-negligible amount, unless
bosonic loops are globally included by the appropriate choice of the
SU(2) gauge coupling. And last not least, first experimental results
from LEP2 gave {\it direct} experimental evidence for
a non-vanishing trilinear coupling among the vector bosons characteristic
for the non-Abelian structure of the $SU(2) \times U(1)$ electroweak
theory.

\vspace{1cm}
\leftline{\bf Acknowledgement}
It is a pleasure to thank Masaaki Kuroda and Ingolf Kuss for collaboration
on the subject matter of the present paper and Stefan Dittmaier for
useful dicussions. The stimulating atmosphere and the magnificent 
hospitality in Ouranoupolis, Greece, at the occasion of the European
network meeting organized by Argyris Nicolaidis, is gratefully 
acknowledged.
\newpage
\leftline{\bf References}
\begin{enumerate}
\item M. Bilenky, J.L. Kneur, F.M. Renard and D. Schildknecht,
Nucl. Phys. {\bf B409} (1993) 22; Nucl. Phys. {\bf B419} (1994) 240.
\item I. Kuss and D. Schildknecht, Phys. Lett. {\bf B383} (1996) 470.
\item The L3 Collaboration, CERN-PPE/97/98 (July 1997), submitted to
Phys. Lett. B
\item J. Maalampi, D. Schildknecht and K.H. Schwarzer, Phys. Lett. 
{\bf B166} (1986) 361;\hfill\break
M. Kuroda, J. Maalampi, K.H. Schwarzer and D. Schildknecht,
Nucl. Phys. {\bf B284} (1987) 271;\hfill\break
M. Kuroda, J. Maalampi, D. Schildknecht and K.H. Schwarzer, Phys. Lett
{\bf B190} (1987) 217.
\item M. Kuroda, I. Kuss and D. Schildknecht, BI-TP 97/15, hep-ph/9705294,
to appear in Phys. Lett. B
\item M. B{\"o}hm, A. Denner and S. Dittmaier, Nucl. Phys. {\bf B376} (1992)
443; err. ibid. {\bf B391} (1993) 483.
\item W. Beenakker et al., in Physics at LEP2, eds. G. Altarelli, T.
Sj{\"o}strand, F. Zwirner, CERN 96-01 Vol. 1, p. 79, hep-ph/9602351.
\item S. Dittmaier, D. Schildknecht and G. Weiglein, Nucl. Phys. {\bf B465}
(1996) 3.
\item H. Burkhardt and B. Pietrzyk, Phys. Lett. B356 (1995) 398;\hfill\break
S. Eidelman and F. Jegerlehner, Z. Phys. {\bf C67} (1995) 585.
\item S. Dittmaier, Talk at the {\it 3rd International Symposium on Radiative
Corrections} Cracow, Poland, August 1996, hep-ph/9610529, Acta Physica
Polonica {\bf B28} (1997) 619.
\item W. Beenakker et al., hep-ph/9612260, NIKHEF 96-031, PSI-PR-96-41.
\end{enumerate}
\end{document}